\documentclass [12pt]{article}
\usepackage{axodraw}

\catcode`@=12
\newcommand{\be}{\begin{equation}}
\newcommand{\ee}{\end{equation}}
\newcommand{\bea}{\begin{eqnarray}}
\newcommand{\eea}{\end{eqnarray}}

\def\Ds{\Delta m_{\odot}^2}
\def\Da{\Delta m_{atm}^2}
\def\s12{\sin\theta_{12}}
\def\s23{\sin\theta_{23}}
\def\s13{\sin\theta_{13}}
\def\t12{\theta_{12}}
\def\t23{\theta_{23}}
\def\t13{\theta_{13}}
\def\e{\epsilon}

\def\cp{\cos\phi}
\def\cps{\cos^2\phi}
\begin{document}
\thispagestyle{empty}
\begin{flushright} UCRHEP-T407\\VEC/Physics/P/1\\March 2006
\end{flushright}
\vspace{0.5in}
\begin{center}
{\LARGE \bf $A_4$ symmetry and  prediction of $U_{e3}$
in a modified Altarelli-Feruglio model\\}
\vspace{1cm}
{\bf Biswajit Adhikary$^{a}$, Biswajoy Brahmachari$^{b}$, 
Ambar Ghosal$^{a}$,\\
Ernest Ma$^{c}$, and M. K. Parida$^{d}$\\}
\vskip .5cm
(a) Saha Institute of Nuclear Physics,\\
1/AF Bidhannagar, Kolkata (Calcutta) 700064, INDIA \\
\vskip .5cm
(b) Department of Physics, 
Vidyasagar Evening College\\
39, Sankar Ghosh Lane, Kolkata 700006, INDIA\\
\vskip .5cm
(c) Physics Department, University of California,\\ 
Riverside, California 92521, USA\\
\vskip .5cm
(d) Department of Physics, Tezpur University, Tezpur 784028, INDIA\\
\vskip 1cm
\end{center}
\begin{center}
\underbar{Abstract} \\
\end{center}
We show that a radiative modification of a recently proposed model by 
Altarelli and Feruglio with softly broken $A_4$ symmetry leads 
naturally to nonvanishing $U_{e3}$ with $\theta_{13}\simeq 2^o - 4^o$. 
The observed mass squared differences for solar and atmospheric neutrinos 
are reproduced, whereas the predicted solar neutrino mixing angle 
is brought down from the tri-bimaximal prediction to be in better 
agreement with the latest global analysis including experimental data 
from KamLAND and SNO.
\newpage
Experimental measurements on  neutrino oscillations
are consistent with nearly maximal atmospheric neutrino  mixing
angle ($\theta_{23}\simeq 45^o$), large but 
less than maximal solar neutrino mixing angle ($\theta_{12}\simeq 34^o$),
and a small ``CHOOZ'' angle ($\theta_{13}< 10^o$).
Whereas the three neutrino 
masses could be quasi-degenerate or hierarchical 
with normal or inverted 
ordering, these values of the mixing angles are 
found to be remarkably close to the conjectured tri-bimaximal mixing 
ansatz of Harrison, Perkins,
and Scott (HPS) \cite{ref1},  
$$
U_{tb}=\pmatrix{\sqrt{2 \over 3} & \sqrt{ 1 \over 3} & 0 \cr
-\sqrt{ 1 \over 6} & \sqrt{ 1 \over 3} & - \sqrt{ 1 \over 2} \cr
-\sqrt{ 1 \over 6} & \sqrt{ 1 \over 3} & \sqrt{ 1 \over 2}
}.
\nonumber
$$ 
The importance of the non-Abelian discrete symmetry group $A_4$ in 
understanding why charged leptons may have very different masses and 
yet a symmetry exists for the neutrino mass matrix has been 
discussed by Ma and collaborators in recent 
papers~\cite{ref2,ref3,ref4}.  In particular, it was shown in \cite{ref4} 
how the HPS ansatz may be realized. In an interesting development, 
Altarelli and Feruglio (AF) have proposed the simplest such model with 
only two parameters in supersymmetric as well as nonsupersymmetric 
cases \cite{ref5}. The relevant $A_4$ symmetric Lagrangian, after 
spontaneous symmetry breaking, becomes 
\begin{eqnarray}
{\cal L}_{\rm{AF}}
 &=& v_d v_T/\Lambda ( y_e e^ce + y_\mu \mu^c\mu + y_\tau\tau^c\tau)
\nonumber\\
&&
 +~x_a v_u^2(u/\Lambda^2)(\nu_e\nu_e + 2 \nu_\mu\nu_\tau)
\nonumber\\
&& +~x_b v_u^2 2v_S/3\Lambda^2(\nu_e\nu_e + \nu_\mu\nu_\mu 
 \nonumber\\
&& +~\nu_\tau\nu_\tau - \nu_e\nu_\mu - \nu_\mu\nu_\tau - \nu_\tau\nu_e) + h.c.
\end{eqnarray}
\noindent
Here $\Lambda$ is the scale of new 
physics ($\equiv$ seesaw scale) and the Higgs content with their 
vacuum expectation values are presented in Table 1 where all 
fields except 
$\chi^+_{\rm i}$ have been used in Ref. \cite{ref5}. 
Then the neutrino 
mass matrix takes the form 
$$
M_\nu^{\rm{AF}} = m_0\pmatrix{a + 2d/3&-d/3&-d/3\cr
                                 -d/3&2d/3&a-d/3\cr
                                 -d/3&a-d/3&2d/3},
\eqno(2)
$$
\noindent
where $a = 2x_a u/\Lambda$, 
$ d= 2x_b v_S/\Lambda$, $m_0=v_{\rm u}^2/\Lambda$.
This is exactly diagonalized \cite{ref4} by the HPS matrix, 
resulting from the underlying $A_4$ symmetry with 

$$
\sin\theta_{12}^0 = \frac{1}{\sqrt{3}},\,\,  
~~\sin\theta_{23}^0 = -\frac{1}{\sqrt{2}},\,\, 
~~\sin\theta_{13}^0 = 0. 
\eqno(3)
$$
\noindent
and mass eigenvalues 
$$
m_1^0 = a+d, ~~m_2^0 = a, ~~m_3^0 = d-a.
\eqno (4).
$$
\noindent 
where the common mass factor $m_0$ has been absorbed into 
$a$ and $d$.


\begin{table}
\begin{center}
\begin{tabular}{|c|c|c|c|}
\hline
{\rm Lepton}& $SU(2)_L$ & $A_4$&\\
\hline
$(\nu_i, l_i)$&2&3&\\
$l_i^c$&1&1&\\
\hline
Scalar&&&{\rm VEV}\\
\hline
$h_u$&2&1&$<h_u^0>$= $v_u$\\
$h_d$&2&1&$<h_d^0>$=$v_d$\\
$\xi$&1&1&$<\xi^0>$ = u\\
$\phi_S$&1&3&$<\phi_S^0>$ = $(v_S,v_S,v_S)$\\
$\phi_T$&1&3&$<\phi_T>$ = $(v_T,0,0)$\\
$\chi_i^+$&1&3&\\
\hline
\end{tabular}
\caption{List of fermion and scalar fields used in this  model.
}
\end{center}
\end{table}

Although this model prediction  of $\theta_{13}^0=0$ is consistent with
the CHOOZ - Palo Verde upper bound, $\sin\theta_{13}^0 <  0.16$
($\theta_{13} <  10^o$) \cite{ref6,ref7}, the actual value of the mixing 
angle is of considerable 
theoretical and experimental 
interest as more accurate values on 
the parameter are  expected to emerge from 
long baseline and future reactor experiments 
such as
Double CHOOZ, Triple CHOOZ,  No$\nu$a and others \cite{ref8}. Thus, it is 
worthwhile to study some modification of the AF model which may predict
$\theta_{13} \neq 0$, which is necessary for  
CP violation in neutrino oscillations. The second observation 
is that the tri-bimaximal mixing matrix  predicts
$\tan^2\theta_{12}^0=0.5$ corresponding to $\theta_{12}^0 = 35.3^o$ whereas 
 a recent global analysis including the KamLAND and the latest SNO data 
gives 
 $(\tan^2\theta_{12})_{expt.}=0.45 \pm 0.05$  corresponding to
 $(\theta_{12}^0)_{expt.} = 33.8^o\pm 1.5^o$ \cite{ref7}.  
Within $1\sigma$ the tri-bimaximal mixing 
prediction just touches the
upper limit of the solar neutrino mixing angle and it is 
 desirable to have a model prediction where the  mixing angle is in 
accord with 
 the central value obtained from global analysis. 

Further, while fitting the available neutrino data the following relations 
between $|a|$ and $|d|$ have been found useful. In the AF parametrisation 
both $\Delta m_{\rm atm}^2$ and $\Ds$ have been fitted with an ansatz,
$$
|d| = -2|a|(1-2R) \cp.
\eqno (5)
$$
\noindent
where $\phi = {\rm arg}(d) - {\rm arg}(a)$ and $R = \Ds/\Da$ 
and this relation manifests a moderate fine tuning of the two parameters. 
On the other hand, one may assume a simpler relationship,
$$
|d| = -2|a| \cp , 
\eqno (6)
$$
in which case $|m_1|=|m_2|$ at the seesaw scale.  In such a scenario, 
while the value of $\Da$ is fitted, $\Ds$ is expected to be generated 
by radiative corrections.  However, such corrections are 
negligible for small neutrino masses $|m_i| << 0.3$~eV in the Standard 
Model (SM) or in the Minimal Supersymmetric Standard Model (MSSM). 
New particles and interactions (i.e. new physics) are then required. 

In the present work we modify the AF model with an aim to accommodate the 
above desirable features with a parametric relation given by
Eq.~(6).  [We note that deviations from tri-bimaximal mixing have 
already been considered in the AF model, using higher-dimensional operators. 
We take the view of starting with the tri-bimaximal form of the 
neutrino mass matrix given by Eq.~(2), however it may arise, and then 
modifying it with a particular radiative mechanism.] 
The model then predicts new values of the mixing 
angles $\theta_{12}$ and $\theta_{13}$ while bringing the former closer to 
the central value of the experimental data and lifting the latter to 
nonvanishing values which are within the accessible limits of planned 
experiments \cite{ref8}.  Because of its structure, any nonvanishing 
value of $\theta_{13}$ in this model is found to be constrained by the 
solar and atmospheric mixing angles, hence we predict only small
values of the CHOOZ angle in the end.

To generate the desirable new radiative contributions to $M_{\nu}$, we 
introduce three singlet charged scalars $\chi^+_i(i =1,2,3)$ 
transforming as an $A_4$-triplet in the nonsupersymmetric version of 
the AF model with two Higgs doublets $h_u$ and $h_d$.  The Lagrangian 
of the present model has three parts, 
$$
 {\cal L} = {\cal L}_{\rm{AF}} + {\cal L}_1 + {\cal L}_2.
\eqno(7) 
$$
\noindent
Here ${\cal L}_{\rm{AF}}$ is already given in Eq.~(1) and in 
Ref.~\cite{ref5}, and ${\cal L}_1$ is the 
additional contribution of the $\chi^+_i$ scalars that respects 
$A_4$ symmetry.  The term ${\cal L}_2$ is introduced to break the 
$A_4$ symmetry softly and in conjunction with ${\cal L}_1$, it gives 
rise to new radiative contributions known often as the Zee 
mechanism \cite{ref9}, as depicted in Fig.~1.  Explicitly ${\cal L}_1$ 
and ${\cal L}_2$ are given by 
$$
{\cal L}_1 = f~(L~L~\chi_i) \subset (3 \times 3 \times 3)
$$
$$
= f(\nu_\mu \tau \chi^+_1+ \nu_\tau e \chi^+_2 + 
\nu_e \mu \chi^+_3
-\nu_\tau \mu \chi^+_1- \nu_e \tau \chi^+_2 
- \nu_\mu e \chi^+_3
).
\eqno(8)
$$
$$
{\cal L}_2 = c_{12}h_u^T i\tau_2 h_d(\chi_1^+ + \chi_2^+ + \chi_3^+).
\eqno(9)
$$
\noindent
It is to be noted that   $ L$ in ${\cal L}_1$ denotes lepton doublets. 
The neutrino mass matrix comes out as 
$$
M_\nu = \pmatrix{a + 2d/3&-d/3&-d/3 - \epsilon\cr
                       -d/3&2d/3&a-d/3 + \epsilon\cr
                       -d/3 - \epsilon&a-d/3 +\epsilon&2d/3},
\eqno(10)
$$
\noindent
where the $a$ and $d$ terms are obtained in 
the same way as in the AF model due to higher dimensional operators. The 
$\epsilon$ terms are the additional contributions from the one-loop 
radiative diagram as shown in Fig.~1,
$$
\epsilon = fm_\tau^2 {\frac{c_{12}v_u}{v_d}}F(m_\chi^2, m_{h_d}^2),
\eqno(11)
$$
\noindent
with the definition, 
$$
F(M_1^2,M_2^2) = {\frac{1}{16\pi^2(M_1^2-M_2^2)}}
\ln{\frac{M_1^2}{M_2^2}}.
\eqno(12)
$$

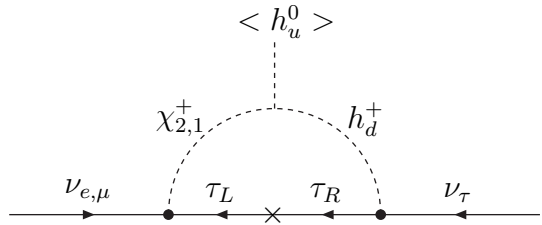
\begin{figure}[htbp]
\begin{center}\begin{picture}(300,100)(0,45)
\ArrowLine(50,50)(110,50)
\ArrowLine(150,50)(110,50)
\ArrowLine(190,50)(150,50)
\ArrowLine(250,50)(190,50)
\Text(80,55)[b]{$\nu_{e,\mu}$}
\Text(220,55)[b]{$\nu_{\tau}$}
\Text(130,55)[b]{$\tau_L$}
\Text(150,50)[]{$\times$}
\Text(170,55)[b]{$\tau_R$}
\Text(115,80)[b]{$\chi^+_{2,1}$}
\Text(185,80)[b]{$h_d^+$}
\DashLine(150,90)(150,115){2}
\Text(155,117)[b]{$<h_u^0>$}
\DashCArc(150,50)(40,0,180){2}\Vertex(110,50){2}\Vertex(190,50){2}
\end{picture}\end{center}
\caption[]{\label{fen}One loop radiative $\nu_{e,\mu}$ - $\nu_\tau$ mass 
due to charged Higgs exchange.}\end{figure}

Apart from the presence of the scalar singlets $\chi_{\rm i}^+$, 
the  effective theory below the seesaw scale in this case is analogous to 
the nonsupersymmetric two-Higgs doublet model (2HDM). 
The masses $m_{\chi_i}$ are degenerate and have been 
generically represented as $m_\chi$.  Furthermore, the interactions 
written in Eqs.~(8) and (9) can generate corrections to all the 
off-diagonal entries of $M_\nu$, but we retain only the dominant terms 
proportional to $m_\tau^2$.  Diagonalizing the neutrino mass 
matrix given in Eq.~(10) with the assumption that $\epsilon$ is small, 
we obtain the three mass eigenvalues as 
$$
m_1 = a+ d + \epsilon,\quad  m_2 = a ,\quad  m_3 = d-a-\epsilon,
\eqno(13)
$$
\noindent
where  the parameters are, in general, complex. Model predictions
for  neutrino oscillations  with $U_{13}=0$ 
have been discussed for 
complex values of $a$ and $d$ in Ref. \cite{ref4,ref5}.
For the sake of simplicity and economy of parameters, we discuss model
predictions by treating all the three parameters to be real and then,
more generally, 
by treating only $d$ as  complex.
\vskip 0.1in
\noindent
{\bf (A) Real parameters}
\vskip 0.1in
\noindent
In this case by treating all the three parameters as real and by 
solving the eigenvalue
equation  the following expressions are obtained in the leading
approximation with $|\e| << |a|,|d|$,
$$
\sin\theta_{12} = {\frac{1}{\sqrt 3}} + \delta_1,\quad 
\sin\theta_{23} = -({\frac{1}{\sqrt 2}} + \delta_2),\quad 
\sin\theta_{13} = \delta_3,
\eqno(14)
$$
\noindent
where
$$
\delta_1 = \frac{\epsilon}{d\sqrt 3},\quad
\delta_2 = \frac{1}{3}\left[{\frac{\epsilon\sqrt 2}{4a}} - 
           \frac{\epsilon}{\sqrt 2 (2a - d)}\right],\quad 
$$
$$  
\delta_3 = \frac{1}{3}\left[{\frac{\epsilon\sqrt 2}{2a}} +
           \frac{\epsilon}{\sqrt 2 (2a - d)}\right].
\eqno(15)
$$
 We find 
that it is 
possible to fit $\Delta m_{\odot}^2 =m_2^2-m_1^2$ and $\Delta m_{\rm atm}^2
=m_3^2-m_2^2$ if the two 
parameters are  related as
$$
 d = -\kappa ~a,
\eqno(16)
$$
\noindent
where  $\kappa$ is a positive rational number. Then 
$m_1 = (1-\kappa)a + \e$, ~$m_2 = a$, ~$m_3 = -(1+\kappa)a - \e$,
$$
 \Delta m_{\rm atm}^2 \simeq (\kappa^2 + 2\kappa)~a^2 ,
\eqno(17)\\
$$
\noindent
$$  
\Delta m_{\odot}^2 = [(2-\kappa)/(2+\kappa)]\Delta m_{\rm atm}^2
 + 2\e (\kappa-1)\sqrt {\frac {\Delta m_{\rm atm}^2} {\kappa^2+2\kappa}},
\eqno(18)
$$
\noindent
where we have used Eq.~(17) to determine the parameter $a$,
$$
 a= \sqrt {\frac {\Delta m_{\rm atm}^2} {\kappa^2+2\kappa}}.
\eqno(19)
$$
\noindent 
In order to estimate the model predictions, we now express
$\e$, mass eigenvalues, and mixing angles in terms of  
$\Delta m_{\rm atm}^2$, $\Delta m_{\odot}^2$, and the positive number 
$\kappa$,
$$
\e= {\left((\kappa^2+2\kappa)\Delta m_{\rm atm}^2\right)^{1/2}\over
{2(\kappa-1)}}\left[\frac{\Delta m_{\odot}^2}{\Delta m_{\rm atm}^2}-
\frac{2-\kappa}{2+\kappa}\right],
\eqno(20) 
$$
$$
m_1 = -(\kappa-1) \sqrt {\frac {\Delta m_{\rm atm}^2} {\kappa^2+2\kappa}}
    + \e,\,\,
m_2 =  \sqrt {\frac {\Delta m_{\rm atm}^2} {\kappa^2+2\kappa}},\,\,
m_3 = -(\kappa+1) \sqrt {\frac {\Delta m_{\rm atm}^2} {\kappa^2+2\kappa}}
    -\e,
\eqno(21)
$$
$$
\sin\theta_{13} = \frac{\kappa(\kappa+3)}{{6\sqrt 2}(\kappa-1)}
\left[\frac{\Delta m_{\odot}^2}
{\Delta m_{\rm atm}^2}-\frac{2-\kappa}{2+\kappa}\right], 
\nonumber
$$
$$
\sin\theta_{12} = {\frac{1}{\sqrt 3}} - \frac{\kappa+ 2}{
{2\sqrt 3}(\kappa-1)}\left[\frac{\Delta m_{\odot}^2}
{\Delta m_{\rm atm}^2}-\frac{2-\kappa}{2+\kappa}\right]\\ 
={\frac{1}{\sqrt 3}} - \frac{\sqrt{6}(\kappa+2)}{\kappa(\kappa+3)}
\sin\theta_{13},
\nonumber
$$
$$ 
\tan^2\theta_{23} = 1 + 
\frac{\kappa^2}{3(\kappa-1)}\left[\frac{\Delta m_{\odot}^2}
{\Delta m_{\rm atm}^2}-\frac{2-\kappa}{2+\kappa}\right]\\
=1 + \frac{{2\sqrt 2}\kappa}{(\kappa+3)}\sin\theta_{13}. 
\eqno(22)
$$ 
\noindent
It is evident from Eqs.~(20) to (22) that $\e$ and corrections to the mixing 
angles depend upon $\kappa$ apart from the experimentally determined 
quantities like $\Delta m_{\odot}^2$, $\Delta m_{\rm atm}^2$, 
and the  ratio, $R = {\Delta m_{\odot}^2}/{\Delta m_{\rm atm}^2}$.
While a positive $\e$ would predict $\sin\theta_{13} > 0$, $\sin\theta_{12} <
1/\sqrt 3$, and $\tan^2\theta_{23} > 1$, a negative value would give
 $\sin\theta_{13} < 0$, $\sin\theta_{12} > 1/\sqrt 3$,~and 
$\tan^2\theta_{23} < 1$. Since the tri-bimaximal prediction 
corresponding to
$\sin\theta_{12} = 1/\sqrt 3$ is 
just on the border line of the maximal value 
allowed by the recent global analysis \cite{ref7}, 
the negative values of $\e$ 
and $\sin\theta_{13}$ which shift $\sin\theta_{12}$ further away are 
strongly disfavored. We thus search for small and positive values of
$\e$ to predict the masses and mixing angles.

We note that the  number $\kappa$ is not arbitrary. It is clear 
from Eq.~(18)
that the smallness of $\Delta m_{\odot}^2$  
compared to $\Delta m_{\rm atm}^2$
requires $\kappa \simeq 2$. For larger values of $\kappa > 2.5$   
 the leading term dominance condition, $|\e| << |a|$, $|d|$, breaks down. 
For values
of $ 2.2 < \kappa < 2.5 $, 
the solar mixing angle prediction falls below the
present $99\%$ confidence limit and results in $\theta_{12}
 < 30^o$. Thus we use the most plausible value $\kappa = 2$ 
(i.e. Eq.~(6) with $\phi = 0$ and corresponding to the symmetry 
limit $|m_1|=|m_2|$) 
and all the parameters in Eqs.~(20) to (22) are determined
in terms of 
$\Delta m_{\odot}^2$, $\Delta m_{\rm atm}^2$ and their ratio 
${R}$ with 
$$
\e = \sqrt 2\sqrt {\Delta m_{\rm atm}^2}{R}, ~~m_1 = -\frac{1}{2\sqrt 2} 
\sqrt {\Delta m_{\rm atm}^2} + 
\sqrt 2\sqrt {\Delta m_{\rm atm}^2}{R},
\nonumber
$$
$$  
m_2 = \frac{1}{2\sqrt 2} \sqrt {\Delta m_{\rm atm}^2}, 
~~m_3 =  -\frac{3}{2\sqrt 2} \sqrt {\Delta m_{\rm atm}^2}-
\sqrt 2\sqrt {\Delta m_{\rm atm}^2}{R},
\eqno(23)
$$
$$
\sin\theta_{13}= \frac{5}{3\sqrt 2}{R}, ~~\sin\theta_{12} = 
\frac{1}{\sqrt 3}-\frac{2\sqrt 6}{5}\sin\theta_{13},
\nonumber 
$$
$$
\tan^2\theta_{23} = 1 +\frac{4\sqrt 2}{5}\sin\theta_{13}.
\eqno(24)
$$
\noindent
 Using the allowed range for $\Delta m_{\odot}^2 = (7.2 - 8.9)\times 10^
{-3}$ eV$^2$,
$\Delta m_{\rm atm}^2 = (1.7 - 3.3)\times 10^{-3}$ eV$^2$, we find $R=
(2.2-5.2)\times 10^{-2}$, and for $\kappa=2$, we obtain
$$
m_1= -0.015 ~{\rm eV}, ~~m_2 = 0.017 ~{\rm eV}, ~~m_3 = -0.055 ~{\rm eV}.
\eqno (25)
$$
\noindent 
Thus the mass eigenvalues are normally ordered \cite{ref4}. 
While there is hierarchy
between $m_{1,2}$ and $m_{3}$ the masses $m_1$ and $m_2$ 
are nearly 
quasi-degenerate.
The kinematical neutrino mass  $|m_{{\nu_e}}|$ and the 
effective neutrino mass $|m_{\rm ee}|$ contributing to neutrinoless double
beta decay are also small,
$$ 
|m_{{\nu_e}}| \simeq |m_{1,2}| \simeq 0.016~{\rm eV},~~~~ |m_{\rm ee}| \simeq
0.01~{\rm eV},
\eqno (26)
$$
\noindent
which are beyond the detection limits of planned experiments in near future
 \cite{ref10,ref11}. The predictions for mixing angles are
$$
\theta_{12} = {31.13}^o - {33.5}^o, ~~ \theta_{13} = {3.5}^o - {1.5}^o,
~~\theta_{23} = 45.5^o-46^o.
\eqno(27)
$$  
\noindent
 In Eq.~(27), the
smaller (larger) value of 
$\theta_{13}$ is correlated with larger (smaller) value
of $\theta_{12}$. Although still larger values of 
$\theta_{13}$ even closer to 
the CHOOZ upper limit are permitted 
by the model they are correlated with smaller
values of $\theta_{12} < 30^o$ and hence are ruled out even at $99\%$ 
confidence level.
For example with $\kappa = 2.25$ we obtain $\theta_{13} = 6^o$, 
but $\theta_{12} =29.2^o$ which is below the range allowed at  $99\%$
level and hence ruled out.
Thus the  prediction of the angle up to $\theta_{13} \simeq 4^o$ is 
quite natural in this model.  The value of $\nu_{\mu}-\nu_{\tau}$ 
mixing angle is
also found to increase
slightly beyond the tri-bimaximal prediction.
\vskip 0.1in
\noindent
{\bf {(B) One complex and two real parameters}}

In this case  we treat $a$ and 
$\epsilon$ to be real but $d$ complex with its 
phase  $\phi(={\rm arg}(d))$. 
In order to maintain the experimentally observed smallness of
$\Delta m_{\odot}^2$ compared to $\Delta m_{\rm atm}^2$ we 
use the relation \cite{ref4}
$$
 |d| = -2a\cos\phi.
\eqno (28)
$$
\noindent
Then  in the leading approximation,
$$
|m_1|^2 = a^2 -2\e a(2\cps-1),~~~ |m_2|^2 = |a|^2,\
\nonumber
$$
$$
|m_3|^2 = (1 + 8\cps)|a|^2 + 2\e a (2\cps-1),
\eqno (29)
$$ 
$$
 \Delta m_{\rm atm}^2 = |m_3|^2 - |m_2|^2 \simeq 8a^2\cps,\
\nonumber
$$
$$
\Delta m_{\odot}^2 = |m_2|^2 - |m_1|^2 = 2\e a (2\cps-1).
\eqno (30) 
$$
\noindent
Eq.~(28) suggests that  $\phi$
lies in the second  quadrant for positive values of $a$.
Eqs.~(28) to (30) give
$$
|a| = -\sqrt {\Delta m_{\rm atm}^2}/(2{\sqrt 2} \cp),
~~~~\left|\frac{\e}{a}\right|= \frac {4\cps}{|2\cps-1|}R,
\eqno (31)
$$
\noindent
Thus, the masses $|m_1|$, $|m_2|$, $|m_3|$, 
and the parameters $|d|$, $|a|$
and $\e$
are  expressed in terms of 
$\Delta m_{\rm atm}^2$, $\Delta m_{\odot}^2$,
${R}$
and
the phase angle $\phi$. For 
example with $\phi = 180^o$, $\Delta m_{\rm atm}^2 
= 2.5\times 10^{-3}$ ~eV$^2$ and $\Delta m_{\odot}^2 
= 8\times 10^{-5}$ ~eV$^2$, $R= 3.2 \times 10^{-2}$ 
and we obtain the same values of 
masses as in Eqs.~(25) and (26) with normal ordering.
With the general expressions for mass eigenvalues given in Eq.~(11) with 
complex $d$, we solve the 
eigenvalue equation and use Eqs.~(28) to (31) to derive the following 
expressions involving the mixing angles,
$$
|\sin\theta_{13}| = \frac{1}{6\sqrt 2}\left|\frac{\e}{a}\right|\left(\frac{9 + 16\cps}{1 + 3\cps}\right)^{1/2}
\nonumber\\
$$
$$
=\frac{\sqrt 2}{3} \left(\frac{9 + 16\cps}{1 + 3\cps}\right)^{1/2}
\frac{\cps}{|2\cps-1|}R,
\eqno (32)
$$
$$
\sin\theta_{12}=\frac{1}{\sqrt 3}- \frac{1}{2\sqrt 3} \left|\frac{\e}{a}\right|
\nonumber\\
$$
$$
= \frac{1}{\sqrt 3}- \frac{2}{\sqrt 3}\frac{\cps}{|2\cps-1|}R,
\eqno (33)
$$
$$
\tan^2\theta_{23} = 1+ \frac{4}{3}\left|\frac{\e}{a}\right|\frac {\cps}{1+3\cps}
\nonumber\\
$$
$$
= 1 + \frac{16}{3}\frac{\cos^4\phi}{|2\cps-1|\left
(1+3\cps\right)}R.
\eqno (34)
$$
\noindent
In the suitable limit of $|\cp| \to 1$, 
Eqs.~(32) to (34) go over , as they
should, to expressions given in  Eq.~(22) 
for the real case with $\kappa = 2$. 
We find that interesting solutions bringing down  the solar neutrino 
mixing from the tri-bimaximal limit with $\theta_{12} < 35.3^o$ while 
increasing $|\sin\theta_{13}|$ substantially from its zero limit are 
possible 
if the phase of the complex parameter $d$ is in the second 
quadrant. 
While $\cp = -1$ gives predictions on mixing 
angles as in Eq.~(27), Eqs.~(32) to (34) can provide substantially 
different values of mixings for certain other values of the parameter, 
$$
\cp = -0.575: ~~\theta_{12}= 33.5^o - 31.2^o,~~\theta_{23} = 45.5^o - 46^o,
 ~~\theta_{13}= 1.7^o - 3.7^o;
\nonumber
$$
$$
\cp = -0.643: ~~\theta_{12}= 31.5^o - 27^o,~~\theta_{23} = 45.7^o - 46.6^o,
 ~~\theta_{13}= 4^o - 8.5^o.
\nonumber
$$
\noindent
Thus, the prediction for CHOOZ angle could be larger than $4^o$ if 
$\theta_{12} < 31.5^o$. For example $\theta_{13}= 5^o$ would require
$\theta_{12} = 30^o$ which is already near the $2.5\sigma$ limit of 
global analysis.
 
\par
The values of light neutrino masses and mixing angles obtained 
by matching the 
experimentally observed values of $\Delta m_{\odot}^2$ and 
$\Delta m_{\rm atm}^2$ are not likely to change significantly
by radiative corrections through renormalization-group (RG) effects 
\cite{ref12}.  This is due to the 
fact that
the mass eigenvalues are small $|m_i|\simeq O(10^{-2})$ eV.
Further $m_1$ and $m_2$ have opposite signs and that prevents 
significant change 
of $\sin\theta_{12}$ by RG effects. 
Although $m_1$ and $m_3$ have the same 
sign, they 
are not so close to produce significant 
changes in the predicted values of
$\sin\theta_{13}$. In the 2HDM, the radiative corrections are
expected to be further reduced in the region of small values 
of $\tan\beta = v_u/v_d \simeq O(1)$.
limit of planned reactor and
long baseline neutrino experiments \cite{ref8}. 
The prediction of solar neutrino mixing angle few 
degrees  below the tri-bimaximal value could be 
tested by precision experiments in near future. 
Prediction of
$\theta_{13} > 5^o$
is possible if the solar neutrino mixing angle is below the 
$2.5\sigma$ limit
of the current global analysis.
We find that the present model successfully explains the 
existing solar, atmospheric and CHOOZ experimental 
results including those from KamLAND and SNO.

The authors acknowledge the hospitality and stimulating environment of 
WHEPP-9 (Ninth Workshop on High Energy Physics Phenomenology)
held at Institute of Physics, Bhubaneswar, India where this study was 
initiated. MKP thanks the Institute of Mathematical 
Sciences, Chennai, for support as Senior Associate.
EM acknowledges support in part by the U.S. Department of Energy under Grant 
No.~DE-FG03-94ER40837.



\end{document}